\documentclass[twocolumn,showpacs,amsmath,floatfix,prl,aps]{revtex4}
\usepackage{graphicx}

\begin{document}
\title{Magnetic properties of LaO$_{1-x}$F$_x$FeAs}
\author{S. Sharma$^{1,2,3}$}
\email{sangeeta.sharma@physik.fu-berlin.de}
\author{J. K. Dewhurst$^{1,2,3}$} 
\author{S. Shallcross$^4$}
\author{C. Bersier$^{1,2,3}$}
\author{F. Cricchio$^5$}
\author{A. Sanna$^{2,6}$}
\author{S. Massidda$^6$}
\author{E. K. U. Gross$^{2,3}$}
\author{L. Nordstr\"om$^5$}
\affiliation{1  Fritz Haber Institute of the Max Planck Society, Faradayweg 4-6, 
D-14195 Berlin, Germany.}
\affiliation{2 Institut f\"{u}r Theoretische Physik, Freie Universit\"at Berlin,
Arnimallee 14, D-14195 Berlin, Germany}
\affiliation{3 European Theoretical Spectroscopy Facility (ETSF)} 
\affiliation{4 Lehrstuhl f\"ur Theoretische Festk\"orperphysik,
Staudstr. 7-B2, 91058 Erlangen, Germany.}
\affiliation{5 Department of Physics, Uppsala University, Box 530, SE-75121 
Uppsala, Sweden.}
\affiliation{6 SLACS-INFM/CNR and Dipartimento di Scienze Fisiche,
Universita' degli Studi di Cagliari, I-09042 Monserrato (CA), Italy}

\begin{abstract}
Using state-of-the-art first-principles calculations we have
elucidated the complex magnetic and structural dependence of LaOFeAs
upon doping. Our
key findings are that (i) doping results in an orthorhombic ground state
and (ii) there is a commensurate to incommensurate
transition in the magnetic structure between $x=0.025$ and
$x=0.04$. Our calculations further imply that in this system magnetic order 
persists up to the onset of superconductivity at the critical doping of $x=0.05$. 
Finally, our investigations
of the undoped parent compound reveal an unusually pronounced dependence of
the magnetic moment on details of the exchange-correlation (xc) functional used
in the calculation. However, for all choices of xc functional 
an orthorhombic structure is found.
\end{abstract}

\pacs{74.25.Jb,67.30.hj,75.30.Fv,75.25.tz,74.25.Kc}
\maketitle

The discovery of high-Tc superconductivity in the iron oxypnictide LaO$_{1-x}$F$_x$FeAs
is attracting a lot of attention, mostly because this class of materials is one 
of the first set of materials to have a high T$_{\rm c}$ like Cuprates
\cite{kamihara,ren}, with a 
striking difference: in oxypnictides the \emph{itinerant magnetic} FeAs layer 
plays a strong role in superconductivity, while in the cuprates this role is 
adopted by highly localized moments in the CuO plane. This of course puts a 
spotlight on the involvement of magnetic 
fluctuations \cite{mazin2,mazin3,giovannetti,opahle}
in the onset of the superconducting state, making iron oxypnictide a playground 
for understanding the mechanism and role of magnetism in high T$_{\rm c}$ 
superconductivity.

Understanding the magnetic ground-state of LaO$_{1-x}$F$_x$FeAs is the 
first crucial step towards uncovering the mystery of the role of magnetism in 
the onset of superconductivity. Determining magnetic ground states using first 
principles density functional theory
(DFT) calculations is a routine procedure. However, in the present case 
this has proved to be a very difficult task.  All experiments,
whether probing long range order via neutron 
scattering \cite{cruz1,luetkens,klauss,nomura}
or local probe studies via the M\"ossbauer effect\cite{klauss}, conclusively 
report an itinerant small (0.25-0.36 $\mu_B$) moment per Fe 
atom \cite{cruz1,luetkens,klauss,nomura}. On the other hand, DFT
calculations (performed using both experimental and optimised atomic
positions) have resulted in moments ranging from 0.47$\mu_B$ to more
than 2.0$\mu_B$. This dramatic spread of results is presumably indicative of 
both a sensitive dependence of the magnetism upon calculational details, as well
as a strong dependence on crystal structure.

In fact, one of the most complex and interesting aspects of magnetism in 
LaO$_{1-x}$F$_x$FeAs is its interplay with structural properties. At a 
temperature of 136 K the undoped
parent compound undergoes a phase transition to a stripe antiferromagnetic (AFM)
order which, additionally, is associated with a structural distortion from
tetragonal to orthorhombic\cite{cruz1,luetkens,nomura}. 
The impact of doping upon structural and magnetic
properties is only beginning to be understood\cite{huang,klauss,luetkens}; 
experiments differ on whether doping (greater than 5\%) suppresses the 
structural phase transition.
Furthermore, a recent experiment finds evidence of incommensurate order upon 
doping\cite{huang}, indicating a rich interaction between structural and 
magnetic order as the parent compound is doped. Highly accurate \emph{ab-initio} 
calculations can play a crucial role in elucidating this
complex behaviour, and in the present work we shall address in detail the
magnetic and structural properties of both doped and undoped LaOFeAs.

In order to keep the numerical analysis as accurate as possible, in the present
work all calculations are performed using the state-of-the-art full-potential
linearized augmented plane wave (FPLAPW)
method \cite{Singh}, implemented within the Elk code \cite{exciting}.
We have taken great care that all relevant calculational parameters are
converged; in particular we use a ${\bf k}$ mesh of
$12\times12\times8$ shifted by [0.5,0.5,0.5], and 188 states per ${\bf k}$-point  
which ensures convergence of the second variational step\cite{Singh}. We
found that using less well converged values of these parameters has
a dramatic impact on the magnetic moment (such computational details can be seen 
in the supplementary materail).

\begin{figure}[ht]
\vspace{1cm}
\centerline{\includegraphics[width=0.8\columnwidth,angle=0]{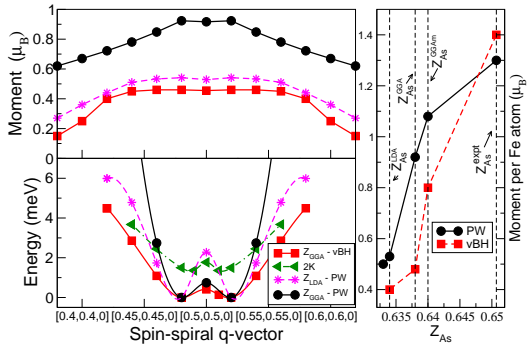}}
\caption{(color online) Top-left hand panel shows the magnetic moment per Fe atom 
(in $\mu_B$) and lower-left hand panel the total energy (in meV) per formula unit. 
All quantities are plotted as a function of spin-spiral $q$-vector.
The right hand panel displays the magnetic moment per Fe atom (in $\mu_B$) as a 
function of position of the As atom in the stripe AFM phase.}
\label{qvsem}
\end{figure}

We first determine the magnetic ground state of the undistorted (tetragonal 
phase) parent compound LaOFeAs.
That magnetism in this system is remarkably sensitive
to calculational details is clear from the spread of moments obtained
using the FP-LAPW method and experimental crystal structure, 0.87-2.20$\mu_B$
\cite{yildirim3,yildirim1,mazin1,anisimov,singh1,yin}.
In this regard it has been noted
that both Fermiology \cite{mazin1} as well as magnetic moment \cite{yin} are 
strongly dependent on the $z$ coordinate of the As atom.
In the right hand panel of Fig.~\ref{qvsem} we show this dependence of the 
magnetic moment on the $z$ coordinate of the As atom for the so-called stripe
AFM phase \cite{yin,yildirim1} (the proposed ground state structure of this 
system), for two common parameterizations\cite{vbh,pw} of the local spin density 
approximation (LSDA). Clearly, this choice leads to a dramatic difference in the 
results.
For instance, at the value of $z_{\rm As}$ obtained by optimizing with the 
generalised gradient approximation (GGA) ($z^{\rm GGA}_{\rm As}=0.638\,$a.u.),
the difference in the moment per Fe atom calculated using two different 
parameterizations for LSDA is 0.44$\mu_B$. 
It is further clear that only for a theoretically optimised $z_{\rm As}$ 
(GGA or LSDA) can a moment close to the experimental value be obtained. Taken 
together with the work of Mazin {\it et al.}, where it was
shown \cite{mazin1} that only for $z^{\rm GGA}_{\rm As}$ a Fermiology
consistent with experiment could be found, these results indicate that
a theoretical value of this parameter should be used. However,
given this unusually strong dependence of magnetism
upon the choice of exchange correlation functional, we adopt
the strategy of calculating structural properties with
the GGA, but then calculate magnetic properties via \emph{both} the von 
Barth-Hedin (vBH)\cite{vbh} and Perdew-Wang (PW)\cite{pw} paramaterizations of 
the LSDA.

Given this, we now explore the phase space of possible magnetic structures
consistent with experiments. To date, the commensurate collinear stripe and 
checker-board 
structures have been investigated, with the stripe phase substantially lower
in energy\cite{dong,yildirim1}.
Two crucial questions which then may be asked are: 
(i) when the constraint of commensuration is removed does the stripe phase
remain the minimal energy structure? and 
(ii) what is the dependence of the moment on the underlying spin structure?

We here calculate spin spirals with ${\bf q}$  around the $[1/2,0,0]$ ($X$)
and $[1/2,1/2,0]$ ($M$) special points in the Brillouin zone.
Two special cases of these are the commensurate stripe and checker-board phases,
generated with wave vectors $M$ and $X$, respectively.
Around the $M$-point, an alternative magnetic order 
is a spin spiral with a $90^{\circ}$ 
phase difference between the two
Fe sites, equivalent to a 2k structure at the $M$-point. Due to the
inherent frustration of ordering at the $M$-point, this structure is also a 
candidate for the ground state \cite{lorenzana}. 
In the upper-left panel of Fig.\ref{qvsem} are shown the moment per Fe atom for 
spin spiral vectors with ${\bf q}$ near the $M$ point for 1k type spin spiral.
Remarkably, we find that for the vBH parameterization a magnetic solution exists 
only around the $M$ point and even a small change in spiral ${\bf q}$ 
from $[0.50,0.50,0.00]$ to $[0.62,0.62,0.00]$ results in a vanishingly small
value of the moment (similar results were obtained using LMTO method in Ref.
\onlinecite{yaresko}). The PW parameterization also
results in a moment which falls stongly on moving away from the $M$ point, but 
the value is non-zero over the whole BZ with a minimal value of 0.20$\mu_B$ at 
the $\Gamma$ point.

Turning to the energetics of these spirals (lower-left panel Fig.\ref{qvsem}), 
we find the 1k type spiral to be always lower in energy than the 2k type spiral
(for clarity we show only the 2k result for vBH-LSDA). 
Interestingly, the global energy minima is not at the $M$ point but instead 
at a spiral of ${\bf q}=[0.52,0.52,0.00]$, a results which is independent of the
functional chosen. The ground state of undistorted LaOFeAs is thus 
an incommensurate spin spiral. With the vBH LSDA this structure lowers 
the total energy by 0.5 meV per formula unit as compared to the previously 
supposed stripe ground state\cite{yin,yildirim1} and gives a moment of 
0.46$\mu_B$. 
This energy difference and the moment are increased to 2 meV per formula unit 
and 0.92$\mu_B$ respectively when the PW LSDA is used for 
determining both the $z_{\rm As}$ and the magnetic structural energies.


\begin{figure}[ht]
\centerline{\includegraphics[width=0.8\columnwidth,angle=0]{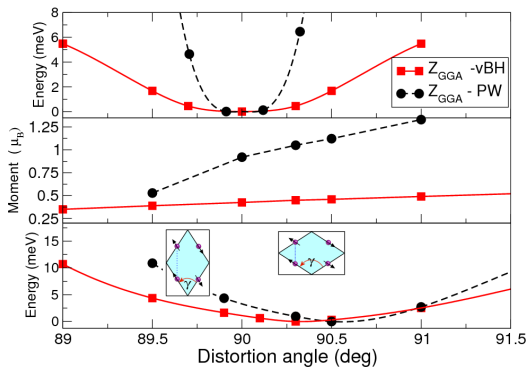}}
\caption{(color online) Shown are the total energies for the non-magnetic system 
(top panel) and the magnetic stripe phase (bottom panel), in meV per formula
unit. In the middle panel is displayed the corresponding moment of the
stripe phase (in $\mu_B$ per Fe atom). All quantities are plotted as a function 
of distortion angle; 90$^\circ$ corresponds to undistorted unit cell. Insets in 
the bottom panel illustrate the distorted magnetic structures.}
\label{dvse}
\end{figure}

At a temperature of $150\,$K LaOFeAs undergoes a phase transition from
the tetragonal to orthorhombic structure, closely associated with the
onset of magnetism\cite{cruz1,luetkens,klauss,nomura,guire,yildirim1,neto}. 
Recently, this transition was explained as the system
lowering magnetic frustration, rationalised by arguments based on a 
$J_1-J_2$ Heisenberg model. Here we shall take a different aproach
and explain the mechanism of this distortion without recourse to the
$J_1-J_2$ model, which is unlikely to be appropriate for a weak
itinerant magnet such as LaOFeAs.

In Fig.~\ref{dvse} (upper panel) is displayed the total energy of the 
non-magnetic state of LaOFeAs as a function of distortion angle, for both
the vBH and PW parameterisations; clearly, a special feature is the very flat 
minima at $90^\circ$. Considering first
the commensurate stripe phase we find that the moment (middle panel)
\emph{increases} with increasing distortion angle. The gain in magnetic energy 
is then sufficient to shift the minimum of the total energy from $90^\circ$
to $90.30^\circ$ (lower panel) for the vBH and $90.50$ for the PW LSDA,
in good agreement with the measured value of $90.27^\circ$.
Thus the vBH LSDA gives \emph{both} the ground state moment and distortion
in close agreement with experiment.
On the other hand, the larger magnetic moment produced by the PW parameterisation
simply results in a slightly larger distortion; reasuringly therefore, the 
qualitative behaviour is independent of the particular choice of functional.

Turning now to possible non-collinearity of 
spins in this distorted phase of LaOFeAs we find that, 
in contrast to the undistorted crystal, the commensurate stripe phase 
is the minimal energy structure in the distorted case (for both parameterizations
of LSDA).
We should stress that the analysis of the structural distortion pesented here
relies solely on the behaviour of the \emph{global moment} with
distortion and the concomitant increase in magnetisation energy, and is thus
appropriate for the itinerant nature of LaOFeAs. In 
Ref.~\onlinecite{yildirim1} the interesting suggestion was made that the
cause of the distortion could be a release of the inherent magnetic frustration
of the FeAs layers. Such a relaxation of frustration would, generally, be
expected to lead to an \emph{increase} in moment, and so this idea is also
compatable with the picture presented here.


\begin{figure}[ht]
\vspace{1cm}
\centerline{\includegraphics[width=0.8\columnwidth,angle=0]{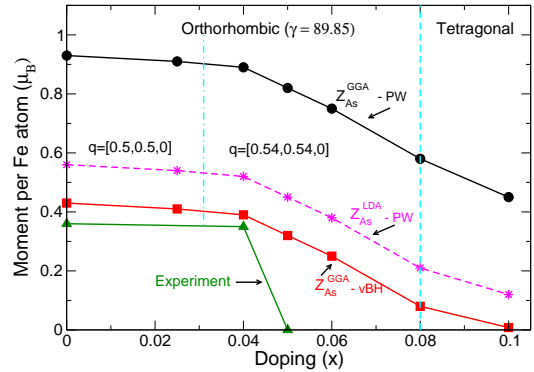}}
\caption{(color online) Spin magnetic moment per Fe atom (in $\mu_B$) as a function 
of electron doping. The experimental data is taken from Ref. \onlinecite{huang}.}
\label{mvsdop}
\end{figure}

The electron doped material LaO$_{1-x}$F$_x$FeAs becomes 
superconducting \cite{kitao,kamihara,huang,luetkens}
at a critical doping of $x=0.05$. A crucial question is whether magnetism
persists up to the superconducting transition, i.e., is
a competing ground state, or is lost before the onset of
superconductivity.
Experimentally, it is established that beyond $x=0.075$ 
no magnetic order persists, however if
magnetic order is entirely lost before the superconducting transition is still 
a point of discussion \cite{huang,luetkens,garcia}. It is also evident from
recent experiments \cite{huang} that there exists a complex structural and
magnetic behaviour with doping, the full nature of which has yet
to be clarified. In fact, on the question of whether the distortion
observed at $x=0$ persists beyond the superconducting transition at $x=0.05$ , 
the current experimental data are contradictory; Refs. \onlinecite{huang} and 
\onlinecite{garcia} find
that distortion persist up to $x=0.08$, while in Refs. \onlinecite{luetkens,klauss} 
no distortion is observed beyond the onset of superconductivity at $x=0.05$.
The magnetic state of of LaO$_{1-x}$F$_x$FeAs is also uncertain, with one 
experiment \cite{huang} finding evidence of an incommensurate structure for 
$x>0$.

\begin{figure}[ht]
\vspace{1cm}
\centerline{\includegraphics[width=0.7\columnwidth,angle=0]{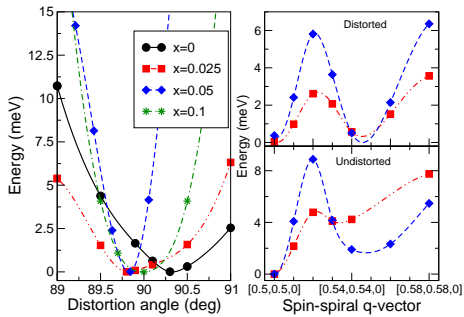}}
\caption{(color online) Left panel shows the total energy (in meV per formula unit) 
as a function of distortion angle, for various dopings. Right panels show  total 
energy (in meV per formula unit) as a function of spin-spiral {\bf q}-vector for 
distorted (upper panel) and undistorted (lower panel) doped LaOFeAs.}
\label{dop}
\end{figure}

In order to clarify this situation we have determined the ground
state for several doping concentrations, by minimising over
both the distortion angle $\gamma$, see Fig. \ref{dop}, and spiral vector 
${\bf q}$ (for 1k type spin configuration).
To simulate doping a small amount of charge is added to the
unit cell along with a compensating background, which ensures charge neutrality.
Since the concentration of electrons is very small this approximation to the 
doping should be good. Indeed, the accuracy of this approximation has
recently been demonstrated\cite{larson}.

We first consider the overall behaviour of the magnetic moment with doping.
This is shown in
Fig.~\ref{mvsdop} for both the vBH and PW parameterisations of the LSDA.
It is immediately apparent that the choice of functional simply leads
to a scaling of the doping curve. However, only for the vBH LSDA do we find
a behaviour that results in quantatitive agreement with available experimental 
data for $x<0.05$, i.e., less than the critical doping.
However, a marked divergence between experiment
and theory occurs after this point; while the theoretical
data show a slowly vanishing tail the experimental
data display a sudden decrease.
Our calculations, therefore, imply that magnetic
and superconducting order are competing ground states in the sense that
magnetism does not die \emph{before} the onset of superconductivity.
The sharp decrease in moment at $x=0.05$ may then be brought
about by the onset of superconductivity. The question of whether
magnetism and superconductivity then coexist for a small doping range,
or if the onset of superconductivity destroys entirely the magentic order,
we cannot answer.
On the other hand, one should note that the small
moment itinerant magnetism of this system implies a strong
role for spin fluctuations \cite{mazin3,opahle,kohama} which is not correctly 
treated by the LSDA functional used in the present calculations. These may act 
to damp the slowly vanishing tail of the moment vs. doping seen in Fig. 
\ref{mvsdop}.

We now turn to the important question of the impact of doping upon the 
crystal and magnetic ground state. Given that the vBH parameterised LSDA
has shown good agreement with experiment for the magnetic moment, distortion
and doping curve, we shall use this functional to explore this question.
We first fix the magnetic structure
to that of the commensurate stripe phase and, for each value of $x$, minimise 
the distortion angle. Remarkably,
we find for all $x>0$ a crystal distortion \emph{opposite} ($\gamma < 90^\circ$)
to that of the undoped parent compound (see Fig. \ref{dop} left panel). Furthermore,
subsequent minimisation of the spin spiral ${\bf q}$ reveals a stripe phase to
incommensurate spin spiral transition (${\bf q} = [0.54,0.54,0]$)
between $x=0.025$ and $x=0.04$ (top right 
panel, Fig. 3). This opposite distortion upon doping can be understood as
a mechanism the system adopts to lower the moment. 
The monotonic increase of the moment with
$\gamma$, displayed in Fig. \ref{dvse} for $x=0$, is also found for all doping
concentrations, and hence a crystal distortion of $\gamma < 90^\circ$
lowers the moment. 

This finding of a commensurate stripe to incommensurate
transition is in agreement with the experimental observation \cite{huang} of 
Huang \emph{et al.}. We should note, however, that the small energy difference 
between the incommensurate and stripe structures (0.4 meV per formula unit)
suggests that this may only be seen at the low temperature (8 K) experiments
performed by
Huang \emph{et al.}. Furthermore, in experiments a small increase in the
parameter $z_{\rm As}^{\rm expt}$ is seen with doping, an effect we have not
included in our calculations. As increasing this parameter results
in an increase of moment, the likely effect of including this would be to shift
the onset of moment lowering mechanisms ($\gamma < 90^\circ$ distortion
and incommensurate order) to somewhat higher values of doping.


To summarize by the means of accurate {\it ab-initio} calculations 
we have given an explanation for the phase transition 
in LaOFeAs based on an itinerant magnetic picture in terms of
an increase in spin polarisation brought about by crystal distortion.
Furthermore, we have elucidated the impact of doping on the ground state;
we find both that the doped material distorts, but with $\theta=89.85$, and 
a stripe phase to incommensurate spin spiral transition takes place
between $x=0.025$ and $x=0.04$. Most importantly, our calculations
indicate that in this system magnetic order persists up to
the onset of superconductivity.

Supplementary material
\begin{figure}[ht]
\vspace{1cm}
\centerline{\includegraphics[width=0.8\columnwidth,angle=0]{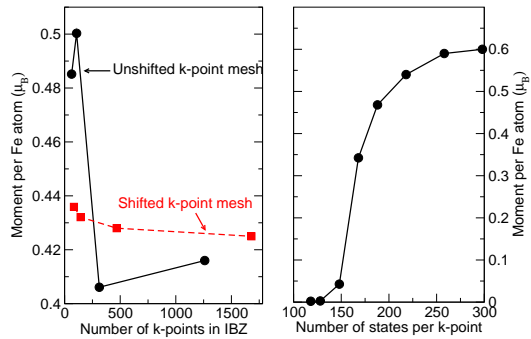}}
\caption{(color online) Left panel shows the change in moment per Fe atom (in
$\mu_B$) as a function of number of ${\bf k}$-points in the IBZ, for these
calculations number of states per ${\bf k}$-point is fixed to 148.
Right panel shows the change in moment per Fe atom (in $\mu_B$) as a function of
the number of states per ${\bf k}$-point, for these calculations number of
${\bf k}$-points is kept fixed at 312 (unshifted mesh).}
\label{s-kpt}
\end{figure}

In the present work all calculations are performed using the state-of-the-art 
full-potential linearized augmented plane wave (FPLAPW) method.
To obtain the Pauli spinor states, the Hamiltonian containing
only the scalar potential is diagonalized in the LAPW basis: this is the
first-variational step. The scalar states thus obtained are then used as
a basis to set up a second-variational Hamiltonian with spinor degrees
of freedom. This is more efficient than simply
using spinor LAPW functions, but care must be taken to ensure that there is
a sufficient number of first-variational eigenstates for convergence of
the second-variational problem.
In Fig. \ref{s-kpt} are shown the results for magnetic moment per Fe atom, for
undistorted undoped LaOFeAs as a function of the number of these
first-variational eigenstates per ${\bf k}$-point. The results are for the
stripe anti-ferromagnetic phase.
It is clear that ground state magnetic moment for this material is greatly
sensitive to both the number of states per  ${\bf k}$-point as well as the
number of ${\bf k}$-points in the irreducible Brillouin-zone (IBZ).

\begin{figure}[ht]
\vspace{1cm}
\centerline{\includegraphics[width=0.8\columnwidth,angle=0]{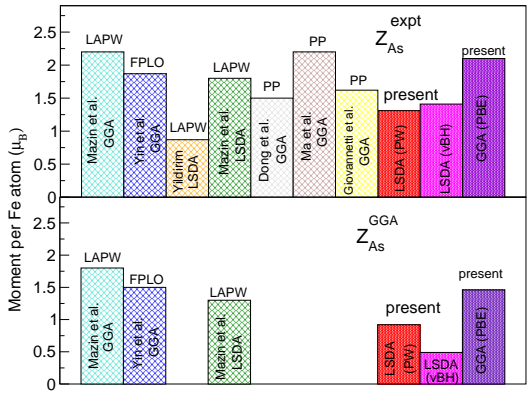}}
\caption{(color online)Top panel shows the magnetic moment per Fe atom
(in $\mu_B$), calculated using various methods, with As atoms at experimental
position. Bottom panel shows the same but for As atom at theoretically optimized
position, this position is obtained using GGA functional and by treating
LaOFeAs non-magnetically.}
\label{s-moms}
\end{figure}

The magnetism in this system is remarkably sensitive
to calculational details is clear from Fig. \ref{s-moms}. In this figure are shown
the results of magnetic moments obtained using various approximations to the 
exchange-correlation functionals implemented within different full-potential as 
well as pseudo potential codes. The moments obtained using different methods and 
experimental As position, range from 0.87-2.20$\mu_B$ and the moments obtained 
using the theoretically optimized (GGA) As position range from 0.46-1.80$\mu_B$. 
On the other hand both Neutron diffraction and M\"ossbauer experiments report
a value between 0.25-0.36$\mu_B$


\end{document}